\PassOptionsToPackage{dvipsnames}{xcolor}
\documentclass[journal, a4paper]{IEEEtran}
\IEEEoverridecommandlockouts
\pdfoutput=1

\renewcommand\IEEEkeywordsname{Index Terms}

\usepackage{graphicx}
\usepackage{caption}
\usepackage[font=footnotesize]{subcaption}
\usepackage{cite}
\usepackage[hyphens]{url}
\usepackage[nolist]{acronym} 
\usepackage{pgfplots}

\usetikzlibrary{arrows,shapes,graphs,graphs.standard,quotes,arrows.meta,decorations.markings,positioning}
\usetikzlibrary{decorations.pathmorphing, shapes.misc}
\pgfplotsset{compat=newest}
\usepackage[bookmarks=false]{hyperref}
\usepackage{amsmath, amsbsy, amssymb}
\usepackage{amsthm}

\usepackage{marginnote}
\tikzset{>=latex}
\definecolor{mittelblau}{RGB}{0, 126, 198}
\definecolor{violettblau}{cmyk}{0.9, 0.6, 0, 0}
\definecolor{rot}{RGB}{238, 28 35}
\definecolor{apfelgruen}{RGB}{140, 198, 62}
\definecolor{gelb}{RGB}{255, 229, 0}
\definecolor{orange}{RGB}{244, 111, 33}
\definecolor{pink}{RGB}{237, 0, 140}
\definecolor{lila}{RGB}{128, 10, 145}
\definecolor{hellgrau}{RGB}{224, 224, 224}
\definecolor{mittelgrau}{RGB}{128, 128, 128}
\definecolor{dunkelgrau}{RGB}{80,80,80}
\definecolor{anthrazit}{RGB}{19, 31, 31}
\definecolor{darkgreen}{RGB}{34,139,34}
\definecolor{aqua}{RGB}{0, 255, 255}

\definecolor{lightgray}{RGB}{211,211,211}

\definecolor{neuesgruen}{RGB}{61, 173, 65}
\definecolor{dunklereshellgrau}{RGB}{176, 176, 176}
\definecolor{neuesgelb}{RGB}{255,160,0}
\definecolor{neuescyan}{RGB}{69,185,224}

\definecolor{tollesgruen}{RGB}{0,217,171}
\definecolor{tollesmagenta}{RGB}{197,67,143}

\definecolor{tollesgelb}{RGB}{255,199,95}
\definecolor{tollesrot}{RGB}{255,111,145}

\tikzset{
       vnd/.style={
        shape=circle,
        fill=black,
        draw,
        inner sep=0pt,
        minimum size=0.2cm},
        cnd/.style={
        shape=rectangle,
        fill=white,
        draw,
        minimum width=0.05mm,
        minimum height = 0.05mm}, 
         vndR/.style={
        shape=circle,
        fill=red,
        draw,
        inner sep=0pt,
        minimum size=0.2cm},
        cndR/.style={
        shape=rectangle,
        fill=white,
        draw=red,
        minimum width=0.05mm,
        minimum height = 0.05mm}
}

\IEEEoverridecommandlockouts

\renewcommand{\vec}[1]{\mathbf{#1}}

\newcommand{\cv}{\vec{c}}

\newcommand{\nv}{\vec{n}}

\newcommand{\sv}{\vec{s}}

\newcommand{\yv}{\vec{y}}

\newcommand{\Hm}{\vec{H}}

\newcommand{\Lm}{\vec{L}}

\begin{document}

\begin{NoHyper}
\title{A Comparative Study of Ensemble Decoding\\Methods for Short Length LDPC Codes}

\author{\IEEEauthorblockN{Felix Krieg, Jannis Clausius, Marvin Geiselhart, and Stephan ten Brink\\}
	\IEEEauthorblockA{
		Institute of Telecommunications, Pfaffenwaldring 47, University of  Stuttgart, 70569 Stuttgart, Germany 
		\\\{krieg,clausius,geiselhart,tenbrink\}@inue.uni-stuttgart.de\\
	}
		\thanks{This work is supported by the German Federal Ministry of Education and Research (BMBF) within the project Open6GHub (grant no. 16KISK019).}
  }

\maketitle

\begin{acronym}
\acro{LBP}{layered BP}
\acro{UER}{undetected error rate}
\acro{ML}{maximum likelihood}
\acro{NED}{noise-aided ensemble decoding}
\acro{SED}{scheduling ensemble decoding}
\acro{BP}{belief propagation}
\acro{BPL}{belief propagation list}
\acro{LDPC}{low-density parity-check}
\acro{BER}{bit error rate}
\acro{SNR}{signal-to-noise-ratio}
\acro{BPSK}{binary phase shift keying}
\acro{AWGN}{additive white Gaussian noise}
\acro{LLR}{Log-likelihood ratio}
\acro{MAP}{maximum a posteriori}
\acro{FER}{frame error rate}
\acro{BLER}{block error rate}
\acro{SCL}{successive cancellation list}
\acro{SC}{successive cancellation}
\acro{BI-DMC}{Binary Input Discrete Memoryless Channel}
\acro{CRC}{cyclic redundancy check}
\acro{CA-SCL}{CRC-aided successive cancellation list}
\acro{PAC}{polarization-adjusted convolutional}
\acro{BEC}{Binary Erasure Channel}
\acro{BSC}{Binary Symmetric Channel}
\acro{BCH}{Bose-Chaudhuri-Hocquenghem}
\acro{RM}{Reed--Muller}
\acro{RS}{Reed-Solomon}
\acro{SISO}{soft-in/soft-out}
\acro{3GPP}{3rd Generation Partnership Project }
\acro{eMBB}{enhanced Mobile Broadband}
\acro{CN}{check node}
\acro{VN}{variable node}
\acro{GenAlg}{Genetic Algorithm}
\acro{CSI}{Channel State Information}
\acro{OSD}{ordered statistic decoding}
\acro{MWPC-BP}{minimum-weight parity-check BP}
\acro{FFG}{Forney-style factor graph}
\acro{MBBP}{multiple-bases belief propagation}
\acro{HRLLC}{hyper-reliable low-latency communications}
\acro{DMC}{discrete memoryless channel}
\acro{SGD}{stochastic gradient descent}
\acro{QC}{quasi-cyclic}
\acro{5G}{fifth generation mobile telecommunication}
\acro{SCAN}{soft cancellation}
\acro{LSB}{least significant bit}
\acro{MSB}{most significant bit}
\acro{AED}{automorphism ensemble decoding}
\acro{EED}{endomorphism ensemble decoding}
\acro{AE-SC}{automorphism ensemble successive cancellation}
\acro{PPV}{Polyanskyi-Poor-Verd\'{u}}
\acro{RREF}{reduced row echelon form}
\acro{PL}{permutation linear}
\acro{LTA}{lower triangular affine}
\acro{BLTA}{block lower triangular affine}
\acro{PTPC}{pre-transformed polar codes}
\acro{MS}{min sum}
\acro{NMS}{normalized min sum}
\acro{OMS}{offset min sum}
\acro{SBP}{saturated belief propagation}
\acro{GED}{gear shift ensemble decoding}
\acro{PG}{projective geometry}
\end{acronym}

\begin{abstract}
To alleviate the suboptimal performance of \ac{BP} decoding of short \ac{LDPC} codes, a plethora of improved decoding algorithms has been proposed over the last two decades.
Many of these methods can be described using the same general framework, which we call \emph{ensemble decoding}:
A set of independent constituent decoders works in parallel on the received sequence, each proposing a codeword candidate. From this list, the \ac{ML} decision is designated as the decoder output.
In this paper, we qualitatively and quantitatively compare different realizations of the ensemble decoder, namely \ac{MBBP}, \ac{AED}, \ac{SED}, \ac{NED} and \ac{SBP}.
While all algorithms can provide gains over traditional \ac{BP} decoding, ensemble methods that exploit the code structure, such as \ac{MBBP} and \ac{AED}, typically show greater performance improvements.
\end{abstract}
\acresetall

\section{Introduction}
The need to meet the requirements of \ac{HRLLC} \cite{ITU2160} renewed the interest in short block-length coding.
As a result, performance limits for the non-asymptotic regime were derived in \cite{polyanski}, and since then the search for codes and decoders that reach the limit has begun.
A survey on this topic was conducted in \cite{coskun}.
With (near) \ac{ML} decoders, convolutional codes and polar codes were shown to perform near the optimum. 
In contrast, a class of codes that has not reached its potential are \ac{LDPC} codes \cite{Gallager} \cite{mackay99bp}. 
Good short-length performance of \ac{LDPC} codes is especially desirable considering efforts to unify channel coding for next-generation communication systems \cite{bits2023unified}.
While other classes are limited by the \ac{ML} performance of the code, \ac{LDPC} decoders usually do not even perform close to the optimum of the code for short block lengths.
In particular, \ac{BP} decoding is sub-optimal due to non-avoidable short cycles in their Tanner graph.
To alleviate this deficit, various improvements over standard \ac{BP} decoding have been proposed.
However, these methods in turn have other drawbacks:
For example, \ac{OSD} \cite{osd} is highly complex and even \ac{OSD} post-processing after \ac{BP} typically exceeds acceptable latency limits.
Methods to avoid trapping sets face the same issues and are specific to a single code \cite{Ryan_TrappingSets}. 

\begin{figure} 
	\centering
	\resizebox{\columnwidth}{!}{\begin{tikzpicture}
\tikzset{
edge/.style = {thick,black},
decrect/.style={rectangle, draw, minimum size=1.2cm, fill=white!90!gray},
intrect/.style={rectangle, draw, minimum size=0.9cm, fill=white!90!gray},
mydiamond/.style={draw, diamond, aspect=2.7,text width=2.0cm, inner sep=0pt,  fill=white!90!red},
}

\tikzstyle{conn} = [-{latex[length=2mm,width=2mm]}];

\node[draw,shape=circle, fill=black, inner sep=0pt,minimum size=0.15cm, label=below:{$\mathbf{L}_\mathrm{ch}$}] (Lch) at (-.5,0.75) {};

\node[decrect] (d1) at (1.5, 3) {$\mathrm{Dec}_1$};
\node[decrect] (d2) at (1.5, 1.0) {$\mathrm{Dec}_2$};
\node[decrect] (d3) at (1.5, -2) {$\mathrm{Dec}_M$};

\coordinate (end) at (4.5,0) {};

\coordinate (help) at (0,4) {};

\draw [edge,conn] (Lch)--(0,0.75)|-(d1);
\draw [edge,conn] (Lch)--(0,0.75)|-(d2);
\draw [edge,conn] (Lch)--(0,0.75)|-(d3);

\node[mydiamond] (cond1) at (5.0, 3.0) {$\hat{\mathbf{c}}_1\cdot \mathbf{H}^\mathsf{T}=\mathbf{0}?$};
\node[mydiamond] (cond2) at (5.0, 1.0) {$\hat{\mathbf{c}}_2\cdot \mathbf{H}^\mathsf{T}=\mathbf{0}?$};
\node[mydiamond] (cond3) at (5.0, -2.0) {$\hat{\mathbf{c}}_M\cdot \mathbf{H}^\mathsf{T}=\mathbf{0}?$};

\draw [edge,conn](d1) to node[below] {$\hat{\cv}_1$} (cond1);
\draw [edge,conn](d2) to node[below] {$\hat{\cv}_2$} (cond2);
\draw [edge,conn](d3) to node[below] {$\hat{\cv}_M$} (cond3);

\draw [edge,conn](cond1) to node[right,red] {\small{no}} (5.0,2);
\draw [edge,conn](cond2) to node[right,red] {\small{no}} (5.0,0);
\draw [edge,conn](cond3) to node[right,red] {\small{no}} (5.0,-3);

\node[rectangle, draw, minimum width=6cm, minimum height=0.3cm, text height=0.3cm, text depth=0.3cm,text centered,rotate=90, fill=white!90!cyan] (decide) at (8, 0.5) {$\hat{\mathbf{c}}=\underset{\hat{\mathbf{c}}_{j},j\in\left\{ 1,\dots,M\right\} }{\mathrm{argmin}}
\operatorname{P}\left(\Lm_\mathrm{ch}|\hat{\cv}_j \right)
$};

\draw [edge,conn](cond1) to node[below] {$\hat{\cv}_1$} (decide.north|-d1.east);
\draw [edge,conn](cond2) to node[below] {$\hat{\cv}_2$} (decide.north|-d2.east);
\draw [edge,conn](cond3) to node[below] {$\hat{\cv}_M$} (decide.north|-d3.east);

\draw [edge,conn] (Lch) -- (0,0.75) -- (help) -- (decide.east|-help) -- (decide.east);

\draw [edge, dotted] (1.5,-0.25)--(1.5,-0.75);
\draw [edge, dotted] (5.0,-0.25)--(5.0,-0.75);
\fill [white] (-0.25,0) rectangle (0.25,-1);
\draw [edge, dotted] (0,-0.25)--(0,-0.75);

\draw [edge,conn](decide) to node[above] {$\hat{\mathbf{c}}$} (9.5,0.5);

\end{tikzpicture}}
	\caption{\footnotesize Block diagram of ensemble decoding  of the channel \acp{LLR} $\mathbf{L}_\mathrm{ch}$ with $M$  constituent hard-output decoders. Only valid codeword candidates are considered in the final \ac{ML}-in-the-list decision.  }
	\label{fig:Block_Diag}
\end{figure}
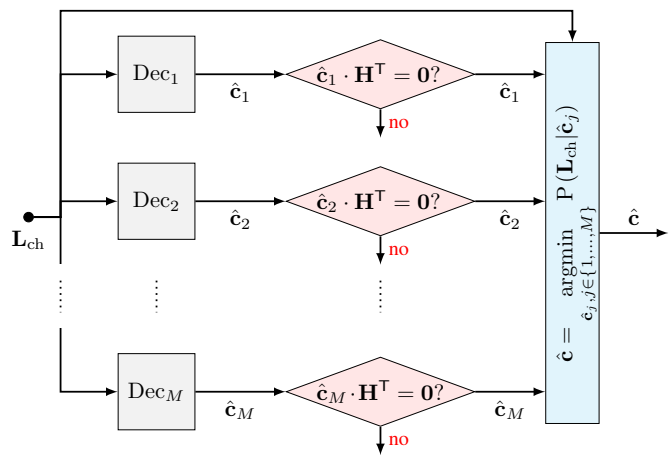

One way forward may be the framework of ensemble decoding, as depicted in Fig.~\ref{fig:Block_Diag} \cite{Hehn_MBBP_cyclic}.
Here, multiple constituent decoders run in parallel and the best codeword candidate is chosen afterwards.
As a result, low latency is ensured, reliability is improved, and error floors are reduced.
Moreover, the independence of each decoder allows for hardware implementations with limited control signal overhead compared to, e.g., \ac{SCL} decoding of polar codes \cite{Kestel2023}.

In this paper, we discuss numerous ensemble schemes for \ac{LDPC} codes, i.e., \ac{MBBP} \cite{Hehn_MBBP_cyclic}, \ac{AED} \cite{ldpcbreakingsymmetries}, \ac{SED}, \ac{NED} \cite{19ned}, and \ac{SBP} \cite{saturatedBP}.
We aim at giving an intuition of the inner workings of each method and discuss their benefits and limitations.
Using Monte-Carlo simulations, we compare their performance for low, medium, and high rate codes.
From this, we can conclude that ensemble decoding benefits enormously from appropriate code design.

Our contributions are outlined as follows:
\begin{itemize}
    \item We give a comprehensive review of ensemble decoding methods, their requirements, and how they are adapted to \ac{BP} decoding of \ac{LDPC} codes.
    \item We provide intuitive explanations for the different mechanisms for constructing decoding ensembles.
    \item Using Monte-Carlo simulations, we quantitatively compare the regarded algorithms in terms of error-rate performance. For this purpose, we introduce the ensemble recovery probability as a measure for the ensemble gain.
\end{itemize}

\section{Preliminaries}

\subsection{LDPC Codes and Belief Propagation Decoding}

\Ac{LDPC} codes are a family of linear codes characterized by their sparse parity-check matrix $\Hm$ \cite{Gallager} \cite{mackay99bp}. 
From the code-defining parity-check matrix, the Tanner graph \cite{tanner} can be build.
This bipartite graph consists of \acp{CN} and \acp{VN} where \ac{CN} $c_i$ is connected to \ac{VN} $v_j$, if $\Hm_{ij}=1$.
Exploiting the sparse connectivity of the graph, the iterative message passing algorithm \ac{BP} can be used to efficiently decode \ac{LDPC} code.
In \ac{BP}, the update equation for the \acp{VN} is 
\begin{equation}\label{eq: vn_update}
    L_{v_i \rightarrow c_j} = L_{\mathrm{ch},i} + \sum_{j'\neq j}L_{c_{j'}\rightarrow v_i},
\end{equation}
where $L_{ch,i}$ is the channel \ac{LLR} and $L_{c_{j'}\rightarrow v_i}$ is the incoming message from \ac{CN} $c_{j'}$.
Similarly, the update for the \acp{CN} is 
\begin{equation}
    L_{c_j \rightarrow v_i} = 2\cdot \tanh^{-1}\left( \prod_{i'\neq i} \tanh \left(\frac{ L_{v_{i'}\rightarrow c_j}}{2} \right) \right).
\end{equation}
After $N_\mathrm{it}$ iterations the final output is calculated by 
\begin{equation}\label{eq: vn_update2}
    L_{v_i} = L_{ch,i} + \sum_{ j}L_{c_{j}\rightarrow v_i}.
\end{equation}

The \ac{BP} algorithm allows for different schedules. 
Here, we cover the flooding and the layered schedule called \ac{LBP}.
A single iteration looks as follows for both cases, respectively.
For the flooding schedule, all \acp{CN} are updated and subsequently all \acp{VN} are updated. 
In contrast, for the layered schedule, only a subset of \acp{CN} $\mathcal{R}_i \subseteq \mathcal{R}$ of size $N_\mathrm{L}$ is updated, followed by the update of their connected \acp{VN}.
Subsequently, the next set $\mathcal{R}_j \subseteq \mathcal{R} \setminus \mathcal{R}_i $ is updated, until all \acp{CN} are updated once.
Typically, layered decoding reduces the number of required decoding iterations by the factor two, however, at an increased latency due to the sequential processing schedule \cite{hocevarlayered}\cite{Litsyn_Schedules_LDPC}.

For long \ac{LDPC} codes, \ac{BP} decoding usually yields near \ac{ML} performance. 
However, this is only guaranteed if the Tanner graph does not contain cycles, or few cycles with large girth.
For short block lengths, this condition is not fulfilled and thus, the performance is compromised.

\subsection{Ensemble Decoding}
Ensemble decoding \cite{Hehn_MBBP_cyclic} helps approaching near \ac{ML} performance with low latency for short block lengths.
Fig.~\ref{fig:Block_Diag} shows the general block diagram of ensemble decoding.
The idea is to have $M$ constituent decoders, working independent and in parallel, leading to possibly different candidates.
The best candidate is chosen according to the \ac{ML}-in-the-list criterion
\begin{align}
    \hat{\mathbf{c}}= \operatorname{Dec}_\mathrm{E}(\Lm_\mathrm{ch}) &=\underset{\hat{\mathbf{c}}_{j},j\in\left\{ 1,\dots,M\right\} }{\mathrm{argmin}}
    \operatorname{P}\left(\Lm_\mathrm{ch}|\hat{\cv}_j \right) \nonumber
    \\
    &=\underset{\hat{\mathbf{c}}_{j},j\in\left\{ 1,\dots,M\right\} }{\mathrm{argmin}}
    \Lm_{\mathrm{ch}}^\mathrm{T}\cdot(-1)^{\hat{\mathbf{c}}_{j}}
    ,
\end{align}
where $\hat{\cv}_j$ is the $j$th candidate and $\Lm_{\mathrm{ch}}$ are the channel \acp{LLR}.
Note that only valid candidates with syndrome $\sv_j=\hat{\cv}_j\Hm^\mathrm{T}=\mathbf{0}$ are considered in the decision process afterwards.
If the constituent decoders output different valid candidates (or convergence behavior), we can observe an ensemble gain.
That means, the ensemble needs diversity in order to work.

\subsection{Automorphisms of (Quasi-) Cyclic Codes}
The automorphism group of a code $\mathcal{C}$ is  $\operatorname{Aut}(\mathcal{C})= \left\{\pi : \pi(\mathbf{c}) \in \mathcal{C} \; \forall \mathbf{c} \in \mathcal{C}\right\}$, meaning the set of permutations that lead from one codeword to another for all codewords \cite{macwilliams77}.
By definition, cyclic and \ac{QC} codes have automorphisms of the form 
\begin{equation}
    \pi_{d,Z}^{\mathrm{QC}}(i)= 
    \begin{cases}
    i+d-Z \quad &\text{if } i \operatorname{mod} Z + d \geq Z\\
    i+d \quad &\text{else},
    \end{cases}
\end{equation}
with $0\leq d < Z$. 
For cyclic codes, we can set $Z=N$ and get the so-called S0 group $\pi_{d,Z}^{\mathrm{S0}}(i)$.
Furthermore, the automorphism group of cyclic codes contain the S1 group \cite[Ch.~8 §5]{macwilliams77} as permutations defined by 
$\pi_d^{\mathrm{S1}} (j) = (j\cdot 2^d)\operatorname{mod} N$ with $d=0,...,N_{\mathrm{S1}}$, where $N_{\mathrm{S1}} = \min \left\{k \,\middle\vert\, k\in\mathbb{N}^+,\, 2^k \equiv 1 \mod N \right\}$.

\section{Ensemble Methods}
\begin{figure*}
    \centering
    \begin{subfigure}[b]{0.4\textwidth}
        \resizebox{\textwidth}{!}{
            \begin{tikzpicture}[>=Latex]

\tikzstyle{codeword}=[circle, fill=black, inner sep=2pt]
\tikzstyle{voronoi}=[line width=2pt]
\tikzstyle{dec1}=[shade,
        top color=mittelblau!60,
        bottom color=mittelblau!30,
        decoration={coil,aspect=-0.25,segment length=3cm,amplitude=.2cm},
        decorate,
        rounded corners=.1cm]
\tikzstyle{dec2}=[shade,
        top color=apfelgruen!60,
        bottom color=apfelgruen!30,
        decoration={coil,aspect=-0.25,segment length=3cm,amplitude=.2cm},
        decorate,
        rounded corners=.1cm]

\tikzset{cross/.style={cross out, draw,line width=2pt,
         minimum size=2*(#1-\pgflinewidth), 
         inner sep=4pt, outer sep=0pt}}

\tikzstyle{message}=[rot, cross]

\draw [dec1]
     (-1.4,-1) -- node [pos=0.3, below, outer sep=0pt] {$\mathrm{Dec}_1$} (1.3,-1) -- (1,0.8) -- (-1,1) -- (-1.4,-1);
\begin{scope}[shift={(5,0)}]
\draw [dec2]
     (-1.4,-1) -- node [pos=0.3, below, outer sep=0pt] {$\mathrm{Dec}_1$} (1.3,-1) -- (1,0.8) -- (-1,1) -- (-1.4,-1);
\end{scope}

\node[message] (y1) at (-0.8,1.5) {};
\node[message] (y2) at (5.8,1.5) {};

\draw[voronoi] (2.5,-1.5) -- (2.5,2.5);

\draw[thick, rot, dashed, ->, line width = 2pt] (y1) edge[bend left=20] node[pos=0.6, above, sloped] {$\pi$} (y2);

\node[codeword] (c0) at (0,0) {};
\node[codeword] (c1) at (5,0) {};

\node[anchor = south east] at (c0) {$\mathbf{c}_0$};
\node[anchor = south] at (c1) {$\mathbf{c}_1=\pi(\mathbf{c}_0)$};

\end{tikzpicture}
        }
        \subcaption{\footnotesize Transformation $\pi$ applied to the received vector $\yv$, as realized by \ac{AED}. The black line marks the Voronoi cell border of the individual codewords.}
        \label{subfig:AED_blob1}
    \end{subfigure}
    \begin{subfigure}[b]{0.19\textwidth}
        \resizebox{\textwidth}{!}{
            \begin{tikzpicture}[>=stealth]

\tikzstyle{codeword}=[circle, fill=black, inner sep=2pt]
\tikzstyle{voronoi}=[line width=2pt]
\tikzstyle{dec1}=[shade,
        top color=mittelblau!60,
        bottom color=mittelblau!30,
        decoration={coil,aspect=-0.25,segment length=3cm,amplitude=.2cm},
        decorate,
        rounded corners=.1cm]
\tikzstyle{dec2}=[shade,
        top color=apfelgruen!60,
        bottom color=apfelgruen!30,
        decoration={coil,aspect=0.25,segment length=3cm,amplitude=-.2cm},
        decorate,
        rounded corners=.1cm]
\tikzstyle{decout}=[dashdotted,
        decoration={coil,aspect=-0.25,segment length=3cm,amplitude=.2cm},
        decorate,
        rounded corners=.1cm]

\tikzset{cross/.style={cross out, draw,line width=2pt,
         minimum size=2*(#1-\pgflinewidth), 
         inner sep=4pt, outer sep=0pt}}

\tikzstyle{message}=[rot, cross]

\draw [dec1]
     (-1.4,-1) -- (1.3,-1) -- (1,0.8) -- (-1,1) -- (-1.4,-1);
\begin{scope}[xscale=-1, yscale=1]
    \draw [dec2]
     (-1.4,-1) -- (1.3,-1) -- (1,0.8) -- (-1,1) -- (-1.4,-1);
\end{scope}
\draw[decout]
     (-1.4,-1) -- (1.3,-1) -- (1,0.8) -- (-1,1) -- (-1.4,-1);

\node [anchor=west, mittelblau] at (1.05,1.7) {$\mathrm{Dec}_1$};
\node [anchor=east, apfelgruen] at (-1.05,1.7) {$\mathrm{Dec}_\pi$};

\draw[dashed, very thick, rot] (0,-1.4) -- node[pos=1.0, right] {$\pi$} (0,2.5);

\node[message] (y1) at (-0.8,1.5) {};

\node[codeword] (c0) at (0,0) {};

\node[anchor = south east] at (c0) {$\mathbf{c}_0$};

\phantom{\node at (0,-1.5){};}

\end{tikzpicture}
        }
        \subcaption{Transformation $\pi$ applied to the decoder, as in \ac{MBBP}, e.g., from automorphisms.}
        \label{subfig:AED_blob_2}
    \end{subfigure}
    \begin{subfigure}[b]{0.19\textwidth}
        \resizebox{\textwidth}{!}{
            \begin{tikzpicture}[>=stealth]

\tikzstyle{codeword}=[circle, fill=black, inner sep=2pt]
\tikzstyle{voronoi}=[line width=2pt]
\tikzstyle{dec1}=[shade,
        top color=mittelblau!60,
        bottom color=mittelblau!30,
        decoration={coil,aspect=-0.25,segment length=3cm,amplitude=.2cm},
        decorate,
        rounded corners=.1cm]
\tikzstyle{dec2}=[shade,
        top color=apfelgruen!60,
        bottom color=apfelgruen!30,
        decoration={coil,aspect=-0.25,segment length=3cm,amplitude=.2cm},
        decorate,
        rounded corners=.1cm]
\tikzstyle{decout}=[dashdotted,
        decoration={coil,aspect=-0.25,segment length=3cm,amplitude=.2cm},
        decorate,
        rounded corners=.1cm]

\tikzset{cross/.style={cross out, draw,line width=2pt,
         minimum size=2*(#1-\pgflinewidth), 
         inner sep=4pt, outer sep=0pt}}

\tikzstyle{message}=[rot, cross]

\draw [dec1]
     (-1.4,-1) -- (1.3,-1) -- (1,0.8) -- (-1,1) -- (-1.4,-1);
\begin{scope}[]
    \draw [dec2]
     (-1.0,-1.2) -- (1.1,-0.6) -- (0.6,0.7) -- (-1.7,1.3) -- (-1.0,-1.2);
\end{scope}
\draw[decout]
     (-1.4,-1) -- (1.3,-1) -- (1,0.8) -- (-1,1) -- (-1.4,-1);

\node [anchor=west, mittelblau] at (1.05,1.7) {$\mathrm{Dec}_1$};
\node [anchor=east, apfelgruen] at (-1.05,1.7) {$\mathrm{Dec}_2$};

\node[message] (y1) at (-0.8,1.5) {};

\node[codeword] (c0) at (0,0) {};

\node[anchor = south east] at (c0) {$\mathbf{c}_0$};

\phantom{\node[outer sep=0pt, inner sep=0pt] at (0,-1.5){};}

\end{tikzpicture}
        }
        \subcaption{Deformation applied to the decoder, as realized by \ac{MBBP} and \ac{SED}.}
        \label{subfig:MBBP_blob}
    \end{subfigure}
    \begin{subfigure}[b]{0.19\textwidth}
        \resizebox{\textwidth}{!}{
            \begin{tikzpicture}[>=Latex]

\tikzstyle{codeword}=[circle, fill=black, inner sep=2pt]
\tikzstyle{voronoi}=[line width=2pt]
\tikzstyle{dec1}=[shade,
        top color=mittelblau!60,
        bottom color=mittelblau!30,
        decoration={coil,aspect=-0.25,segment length=3cm,amplitude=.2cm},
        decorate,
        rounded corners=.1cm]
\tikzstyle{dec2}=[shade,
        top color=apfelgruen!60,
        bottom color=apfelgruen!30,
        decoration={coil,aspect=-0.25,segment length=3cm,amplitude=.2cm},
        decorate,
        rounded corners=.1cm]
\tikzstyle{decout}=[dashdotted,
        decoration={coil,aspect=-0.25,segment length=3cm,amplitude=.2cm},
        decorate,
        rounded corners=.1cm]

\tikzset{cross/.style={cross out, draw,line width=2pt,
         minimum size=2*(#1-\pgflinewidth), 
         inner sep=4pt, outer sep=0pt}}

\tikzstyle{message}=[rot, cross]

\draw [dec1]
     (-1.4,-1) -- (1.3,-1) -- (1,0.8) -- (-1,1) -- (-1.4,-1);

\node [anchor=west, mittelblau] at (1.05,1.7) {$\mathrm{Dec}_1$};

\node[message] (y1) at (-0.8,1.5) {};

\node[message] (y2) at (-1.1,0.4) {};

\draw[very thick, dashed, ->, rot] (y1) -- (y2);

\node[codeword] (c0) at (0,0) {};

\node[anchor = south east] at (c0) {$\mathbf{c}_0$};

\phantom{\node[outer sep=0pt, inner sep=0pt] at (0,-1.5){};}

\end{tikzpicture}
        }
        \subcaption{Translation of received vector $\yv$, as realized by \ac{NED} and \ac{SBP}.}
        \label{subfig:Noise_blob}
    \end{subfigure}
    
    \caption{\footnotesize Illustration of how the decoding regions interact with the received vector ({\color{rot}$\boldsymbol{\times}$}) for different ensemble decoding variants. In each scenario, the initial decoder $\operatorname{Dec}_1$ fails, but the ensemble of two decoders can successfully decode.}
    \label{fig:blobs}
\end{figure*}
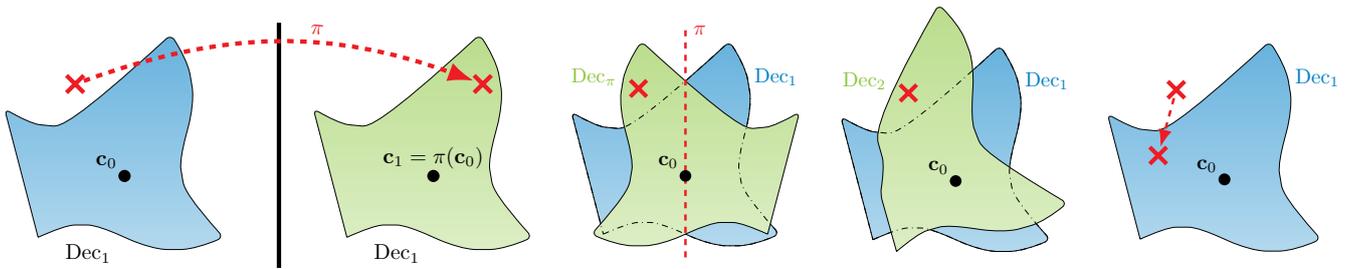

\subsection{Multiple Bases Belief Propagation (MBBP)}
For \ac{MBBP} \cite{Hehn_MBBP_cyclic}, the ensemble is generated from \ac{BP} decoding based on different parity-check matrices $\Hm_j$, thus, $\hat{c}_j = \operatorname{Dec}_{\Hm_j}(\mathbf{\Lm_\mathrm{ch}})$. 
Note that each decoder still operates on a valid parity-check matrix for the code. 
The different matrices should be constructed from low-weight parity-checks (i.e., low-weight codewords of the dual code). 
As a result, the constructed matrix is sparse and suitable for \ac{BP} decoding.
In general, however, these low-weight codewords are hard to find. 
Some code families, such as cyclic codes simplify this task, as each parity-check matrix can be constructed from a single low weight parity-check and its cyclic shifts.
In general, the quality of the candidates might vary, since some matrices are better suited than others for \ac{BP}.

\subsection{Automorphism Ensemble Decoding (AED)}
\Ac{AED} \cite{rm_automorphism_ensemble_decoding} exploits symmetries in the code, allowing to feed the decoder permuted variants of the observation $\yv$.
In other words, each decoder perceives $\yv$ from a different perspective, possibly leading to different candidates, as long as the decoder is not invariant to this transformation.
This invariance is called the symmetry of the decoder.
Thus, the code needs to have more symmetries than the decoder for \ac{AED} to work. 
Each constituent decoder pre-processes $\Lm_\mathrm{ch}$ with a different permutation and decodes it.
After decoding, the permutation is reversed and, thus, a candidate is calculated by $\hat{c}_j = \pi^{-1}_j\left(\operatorname{Dec}(\pi_j(\mathbf{\Lm_\mathrm{ch}}))\right)$.

While being attractive due to their known automorphism group, \ac{AED} cannot be directly applied to \ac{QC} \ac{LDPC} codes.
This is because the circulant structure of the parity-check matrix imposes an invariance of \ac{BP} decoding with respect to \ac{QC} permutations
\cite{Chen_Cyclic_LDPC_AED, ldpcbreakingsymmetries}.
Therefore, \cite{ldpcbreakingsymmetries} proposes to break the symmetry of the parity-check matrix in either of three ways, namely the deletion of rows, the addition of rows, or the summation of rows. If the code contains more symmetries, e.g., is cyclic, \ac{AED} can be directly applied \cite{Chen_Cyclic_LDPC_AED}.

Instead of permuting $\yv$, applying the automorphism to the decoder leads to an equivalent decoding behavior. This form of \ac{AED} is a special case of \ac{MBBP}, where the parity-check matrices are column permutations of each other.
The advantage of \ac{AED} over general \ac{MBBP} is that on average, all candidates perform equally well, with similar latency.

Very recently, two generalizations of \ac{AED} have been proposed.
The first, called generalized \ac{AED}, considers code automorphisms that are arbitrary linear, bijective maps, rather than just permutations \cite{mandelbaum2023gaed}.
The second, namely \ac{EED}, further generalizes this to transformations that are not even bijective \cite{mandelbaum2024eed}.
Both decoding algorithms, however, require specially tailored codes or only work for very short codes, so they are not yet universally applicable, and thus, are not considered in this paper.

\subsection{Scheduling Ensemble Decoding (SED)}
\Ac{SED} is based on different schedules for the \ac{CN} processing. 
Neither the inputs nor the outputs of the constituent decoders are altered. 
The diversity comes from the order of processing of the the \ac{CN} subsets $\mathcal{R}_i$ with $ i=0,...,\frac{N}{N_\mathrm{L}}-1$.
The number of different schedules is $\frac{N}{N_\mathrm{L}}!$.
The parameter $N_\mathrm{L}$ dictates the level of parallelism in the decoder.
Considering contention free memory access, the maximum for \ac{QC} \ac{LDPC} codes is $N_\mathrm{L}=Z$.
For non-\ac{QC} codes, the maximum is non-trivial.
Note that the performance per branch might vary, i.e., some schedules may be inherently better suited than others.
In particular, for the 5G \ac{LDPC} code, a na\"ive top-to-bottom schedule of the rows is highly suboptimal in performance. 
For polar codes, \ac{BP} scheduling ensembles are proposed in \cite{CRC_BPL_ISIT20}. However, to the best of our knowledge, \ac{SED} for \ac{LDPC} codes has not yet been published in literature.\footnote{This idea originated from a discussion with Tom Richardson at the Munich Workshop on Shannon Coding Techniques 2024.}

\subsection{Noise-aided Ensemble Decoding (NED)}
For \ac{NED}, identical constituent \ac{BP} decoders based on the unaltered parity-check matrix are used.
The ensemble diversity comes from an alteration of the channel \acp{LLR} $\Lm_\mathrm{ch}$.
For all constituent decoder but the first one, Gaussian noise is added as $\Lm_{j,\mathrm{ch}} = \Lm_\mathrm{ch}+\nv_j$, where $\nv \sim \mathcal{N}(0,\sigma_\mathrm{NED}^2)$.
Note that $\sigma_\mathrm{NED}^2$ is a hyper parameter and must be optimized.
Furthermore, adding noise leads to a degradation in performance per constituent decoder, which is, however, almost negligible.

\subsection{Saturated Belief Propagation (SBP)}
While originally proposed as an ``afterburner'' that is only applied when conventional \ac{BP} decoding fails \cite{saturatedBP}, \ac{SBP} can be applied directly as an ensemble decoder. 
Similarly to \ac{NED}, \ac{SBP} alters the decoder inputs rather than the constituent decoders themselves.
Here, the $\log_2(M)$ most unreliable positions of $\Lm_\mathrm{ch}$ are saturated (set to $\pm \infty$), according to all $M$ possibilities.
These are then the input to the different branches in the ensemble, where the decoding continues.

\section{Intuitions of Ensemble Decoding}

To give an intuition on the workings of the different ensemble decoding methods, Fig.~\ref{fig:blobs} visualizes the effects of changing the constituent decoders or their inputs on the decoding outcome.
The advantage of an ensemble decoder arises from the varied decoding patterns exhibited by its sub-optimal constituent decoders. Each of these imperfect (non-ML) decoders has a unique decoding region that does not align with the Voronoi cell (i.e., the decoding region of an \ac{ML} decoder) of a codeword.
Within this abstract region, the decoder is assumed to successfully converge to its associated codeword.
In particular, for \ac{BP}-based decoders, the majority of errors stems from the algorithm not converging at all, rather than to a erroneous codeword.
In Appendix \ref{sec:appendix}, we demonstrate this property by Monte-Carlo simulation.
Therefore, the decoding region is typically significantly smaller than the Voronoi cell. 
It is important to note that these decoding regions can be highly asymmetrical and even discontinuous.
By applying the \ac{ML} metric to the resulting candidate list of the ensemble, the decoding region of the ensemble decoder can be interpreted as the union of the individual decoding regions. 
If the received vector $\yv$ (marked by {\color{rot}$\boldsymbol{\times}$} in Fig.~\ref{fig:blobs}) falls within the decoding region of any of the constituent decoders, the ensemble is considered to have successfully estimated the correct codeword.
Whereas in general a decoder with a smaller decoding region would result in worse \ac{BLER} performance, even a ``bad'' decoder can be beneficial to an ensemble, as long as it covers a region that is not already covered. 
Concluding, the larger the area covered by the ensemble compared to any single decoder, the larger the diversity.

\autoref{subfig:AED_blob1} illustrates an ensemble of size two and the effect of a transformation of a codeword, e.g., a permutation ($\pi: \mathcal{C} \to \mathcal{C}$) of a codeword $\mathbf{c}_1$ onto another codeword $\mathbf{c}_2 = \pi(\mathbf{c}_1)$. Due to the asymmetrical decoding regions of the constituent decoder, the noise pattern around $\mathbf{c}_1$ leads to no convergence. However, the transformed channel observation is correctable by the other decoder. 
Note that the second decoder can only decode this observation because its decoding region is not invariant under the transformation $\pi$. 

The same ensemble can also be constructed by applying the transformation directly to the decoder.
For example, \ac{AED} can be realized as \ac{MBBP} by permuting the columns of the parity-check matrix according to the automorphisms $\pi$ (rather than the observation $\yv$).
\autoref{subfig:AED_blob_2} illustrates the decoding regions of both the original decoder and the transformed decoder. The decoding region of the ensemble decoder is the union of these two regions.

Similarly, \autoref{subfig:MBBP_blob} shows the deformation of a decoding region for two different decoders, as realized by general \ac{MBBP} or \ac{SED}.
Since the second decoder ({\color{apfelgruen!85!black}$\mathrm{Dec}_2$}) covers areas where the initial decoder would fail, the ensemble gains in performance.

Lastly, translation-based decoding schemes such as \ac{NED} and \ac{SBP} are presented in \autoref{subfig:Noise_blob}. They also provide the ensemble with additional gain, as they may nudge $\yv$ into a decoding region.

\section{Results}
\subsection{Code Parameters}
\begin{table}[b]
    \caption{\footnotesize Considered \ac{LDPC}(-like) codes and their parameters}
    \centering
    \begin{tabular}{c|cccc}
        Code & $N$ & $K$ & $R$ & Automorphisms \\
        \hline
        Simplex & 63 & 6 & $0.1$ & general linear, cyclic \\
        5G LDPC & 132 & 66 & $0.5$ & quasi-cyclic \\
        PG LDPC & 273 & 191 & $0.7$ & cyclic \\
    \end{tabular}
    \label{tab:codes}
\end{table}
To compare the different ensemble decoding algorithms, we consider a variety of short \ac{LDPC} and \ac{LDPC}-like codes, listed in Tab.~\ref{tab:codes}.
The first code is a short, low-rate $(63,6)$ simplex code, i.e., the dual of the $(63,57)$ Hamming code.
Due to this property, it is easy to find suitable low-weight parity-checks and also, its automorphism group is large.
The second code is the $(132,66)$ 5G \ac{LDPC} code. It is quasi cyclic with $Z=11$.
Finally, we consider the relatively high-rate, cyclic $(273,191)$ \ac{LDPC} code constructed from a \ac{PG} \cite{LinFGLDPC}. Due to its construction, there are $273$ minimum weight parity-checks, from which suitable subsets can be selected for decoding.
For a fair comparison in terms of complexity, we restrict decoders to have full-rank parity-check matrices with the same number of checks, i.e., we do not consider overcomplete parity-check matrices. An exception is \ac{AED} for the 5G \ac{LDPC} code, which uses an undercomplete $\Hm$ matrix obtained from removing the first three parity-checks to break the symmetries of the decoder \cite{ldpcbreakingsymmetries}.

All decoders use the same ensemble size of $M=8$ with each decoder using $8$ flooding or $4$ layered iterations, respectively.
For the \ac{NED} we optimized the noise variance, resulting in $\sigma^2_\mathrm{NED}=0.64$ for $M=4$ and $\sigma^2_\mathrm{NED}=0.25$ for $M=8$.
For the layered \ac{BP} decoder without an ensemble, we chose a top-to-bottom schedule of the check node sets.
Further, we set $N_\mathrm{L}=1$ for the cyclic codes and $N_\mathrm{L}=Z=11$ for the 5G \ac{LDPC} code.
The schedules in the \ac{SED} are chosen randomly and are not optimized.

\subsection{Error-Rate Performance}
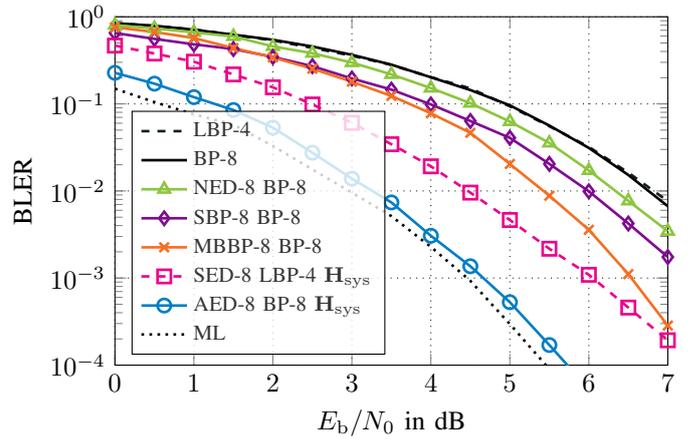
\begin{figure}
    \centering
    \begin{tikzpicture}
\begin{axis}[
width=\linewidth,
height=.7\linewidth,
grid style={dotted,anthrazit},
xmajorgrids,
yminorticks=true,
ymajorgrids,
legend columns=1,
legend pos=south west,   
legend cell align={left},
legend style={fill,fill opacity=0.8},
xlabel={$E_\mathrm{b}/N_0$ in dB},
ylabel={BLER},
legend image post style={mark indices={}},%
ymode=log,
mark size=2.5pt,
xmin=0,
xmax=7,
ymin=1e-4,
ymax=1
]

\addplot[anthrazit , dashed, mark options={solid},mark=none, line width=1.0pt, mark size= 2pt] 
table[col sep=comma]{
0.00, 8.479e-01
0.50, 7.836e-01
1.00, 7.151e-01
1.50, 6.207e-01
2.00, 5.357e-01
2.50, 4.391e-01
3.00, 3.576e-01
3.50, 2.801e-01
4.00, 2.028e-01
4.50, 1.507e-01
5.00, 9.371e-02
5.50, 5.539e-02
6.00, 3.207e-02
6.50, 1.623e-02
7.00, 7.639e-03
};
\label{plot:simplex_layered}
\addlegendentry{\footnotesize LBP-4};

\addplot[color=black,line width = 1pt, solid, mark=none, mark options={solid}]
table[col sep=comma]{
0.00, 8.441e-01
0.50, 7.898e-01
1.00, 7.101e-01
1.50, 6.234e-01
2.00, 5.444e-01
2.50, 4.558e-01
3.00, 3.661e-01
3.50, 2.821e-01
4.00, 2.021e-01
4.50, 1.442e-01
5.00, 9.620e-02
5.50, 5.632e-02
6.00, 3.120e-02
6.50, 1.526e-02
7.00, 6.681e-03
7.50, 2.460e-03
8.00, 8.485e-04
8.50, 2.938e-04
9.00, 6.690e-05
};
\label{plot:simplexrbp8}
\addlegendentry{\footnotesize BP-8};

\addplot[apfelgruen, mark options={solid},mark=triangle, line width=1.0pt] 
table[col sep=comma]{
0.0,0.802
0.5,0.735
1.0,0.663
1.5,0.594
2.0,0.457
2.5,0.382
3.0,0.298
3.5,0.218
4.0,0.152
4.5,0.102
5.0,0.0627
5.5,0.0356
6.0,0.0172
6.5,0.0077
7.0,0.00345
};
\label{plot:simplex_NED_8}
\addlegendentry{\footnotesize NED-8 BP-8};	

\addplot[color=lila,line width = 1pt, solid, mark=diamond, mark options={solid}]
table[col sep=comma]{
0.00, 6.514e-01
0.50, 5.580e-01
1.00, 4.830e-01
1.50, 4.266e-01
2.00, 3.457e-01
2.50, 2.710e-01
3.00, 1.957e-01
3.50, 1.459e-01
4.00, 9.859e-02
4.50, 6.360e-02
5.00, 4.071e-02
5.50, 2.045e-02
6.00, 9.901e-03
6.50, 4.229e-03
7.00, 1.743e-03
};
\label{plot:simplexsbp8}
\addlegendentry{\footnotesize SBP-8 BP-8};

\addplot[color=orange,line width = 1pt, solid, mark=x, mark options={solid}]
table[col sep=comma]{
0.00, 7.581e-01
0.50, 6.684e-01
1.00, 5.702e-01
1.50, 4.379e-01
2.00, 3.419e-01
2.50, 2.533e-01
3.00, 1.796e-01
3.50, 1.234e-01
4.00, 7.801e-02
4.50, 4.655e-02
5.00, 2.043e-02
5.50, 8.814e-03
6.00, 3.581e-03
6.50, 1.110e-03
7.00, 2.847e-04
};
\label{plot:simplexmbbp8}
\addlegendentry{\footnotesize MBBP-8 BP-8};

\addplot[magenta, dashed, mark options={solid},mark=square, line width=1.0pt] 
table[col sep=comma]{
0.0, 0.467000
0.5, 0.379000
1.0, 0.305000
1.5, 0.218000
2.0, 0.155000
2.5, 0.098500
3.0, 0.060600
3.5, 0.034400
4.0, 0.019200
4.5, 0.009570
5.0, 0.004640
5.5, 0.002170
6.0, 0.001090
6.5, 0.000458
7.0, 0.000193
};
\label{plot:simplex_SED_8}
\addlegendentry{\footnotesize SED-8 LBP-4 $\Hm_\mathrm{sys}$};

\addplot[color=mittelblau,line width = 1pt, solid, mark=o, mark options={solid}]
table[col sep=comma]{
0.00, 2.274e-01
0.50, 1.707e-01
1.00, 1.195e-01
1.50, 8.503e-02
2.00, 5.334e-02
2.50, 2.749e-02
3.00, 1.385e-02
3.50, 7.382e-03
4.00, 3.059e-03
4.50, 1.366e-03
5.00, 5.278e-04
5.50, 1.701e-04
6.00, 5.322e-05
};
\label{plot:simplexaeds08}
\addlegendentry{\footnotesize AED-8 BP-8 $\Hm_\mathrm{sys}$};

\addplot[color=black,line width = 1pt, dotted, mark=none, mark options={solid}]
table[col sep=comma]{
0.00, 1.494e-01
0.50, 1.049e-01
1.00, 7.583e-02
1.50, 5.788e-02
2.00, 3.285e-02
2.50, 1.767e-02
3.00, 9.576e-03
3.50, 5.143e-03
4.00, 2.281e-03
4.50, 9.288e-04
5.00, 2.963e-04
5.50, 8.993e-05
6.00, 2.359e-05
};
\label{plot:simplexml}
\addlegendentry{\footnotesize ML};

\end{axis}

\end{tikzpicture}
    \caption{\footnotesize \Ac{BLER} performance comparison for the $(63,6)$ simplex code. $\Hm_\mathrm{sys}$ indicates the use of the systematic parity-check matrix.}
    \label{fig:63_cw}
\end{figure}

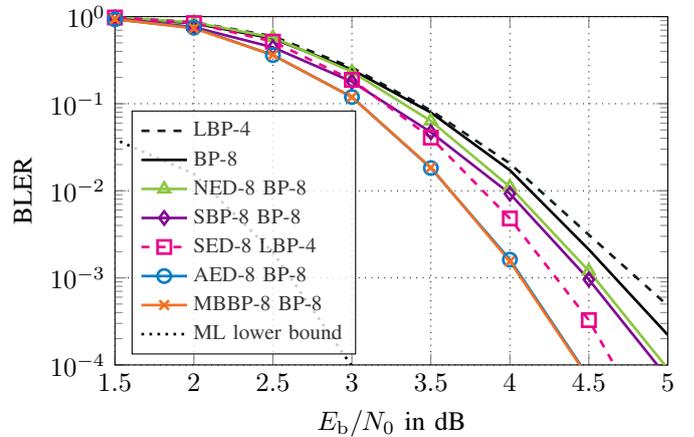
\begin{figure}
    \centering
    \begin{tikzpicture}
\begin{axis}[
width=\linewidth,
height=.7\linewidth,
grid style={dotted,anthrazit},
xmajorgrids,
yminorticks=true,
ymajorgrids,
legend columns=1,
legend pos=south west,   
legend cell align={left},
legend style={fill,fill opacity=0.8},
xlabel={$E_\mathrm{b}/N_0$ in dB},
ylabel={BLER},
legend image post style={mark indices={}},%
ymode=log,
mark size=2.5pt,
xmin=1.5,
xmax=5,
ymin=1e-4,
ymax=1
]

\addplot[anthrazit , dashed, mark options={solid},mark=none, line width=1.0pt, mark size= 2pt] 
table[col sep=comma]{
0.0,1.0
0.5,0.999
1.0,0.999
1.5,0.982
2.0,0.864
2.5,0.595
3.0,0.261
3.5,0.0838
4.0,0.0206
4.5,0.00312
5.0,0.00048
};
\label{plot:273_layered}
\addlegendentry{\footnotesize LBP-4};

\addplot[color=black,line width = 1pt, solid, mark=none, mark options={solid}]
table[col sep=comma]{
0.00, 1.000e+00
0.50, 1.000e+00
1.00, 1.000e+00
1.50, 9.738e-01
2.00, 8.382e-01
2.50, 5.599e-01
3.00, 2.466e-01
3.50, 7.919e-02
4.00, 1.721e-02
4.50, 2.115e-03
5.00, 2.209e-04
};
\label{plot:273rbp8}
\addlegendentry{\footnotesize BP-8};

\addplot[apfelgruen , mark options={solid},mark=triangle, line width=1.0pt] 
table[col sep=comma]{
0.0,1.0
0.5,1.0
1.0,0.998
1.5,0.979
2.0,0.858
2.5,0.577
3.0,0.233
3.5,0.0634
4.0,0.0111
4.5,0.0012
5.0,9.03e-05
};
\label{plot:273_JC_ned}
\addlegendentry{\footnotesize NED-8 BP-8};

\addplot[color=lila,line width = 1pt, solid, mark=diamond, mark options={solid}]
table[col sep=comma]{
0.00, 1.000e+00
0.50, 1.000e+00
1.00, 9.886e-01
1.50, 9.559e-01
2.00, 7.566e-01
2.50, 4.439e-01
3.00, 1.791e-01
3.50, 4.699e-02
4.00, 9.264e-03
4.50, 9.565e-04
5.00, 7.064e-05
};
\label{plot:273sbp8}
\addlegendentry{\footnotesize SBP-8 BP-8};

\addplot[magenta, dashed , mark options={solid},mark=square, line width=1.0pt] 
table[col sep=comma]{
0.0,1.0
0.5,1.0
1.0,0.997
1.5,0.978
2.0,0.843
2.5,0.516
3.0,0.188
3.5,0.0408
4.0,0.00481
4.5,0.000327
5.0,8.2e-06
};
\label{plot:273_JC_SED}
\addlegendentry{\footnotesize SED-8 LBP-4};

\addplot[color=mittelblau,line width = 1pt, solid, mark=o, mark options={solid}]
table[col sep=comma]{
0.00, 1.000e+00
0.50, 1.000e+00
1.00, 9.924e-01
1.50, 9.317e-01
2.00, 7.464e-01
2.50, 3.614e-01
3.00, 1.194e-01
3.50, 1.823e-02
4.00, 1.619e-03
4.50, 7.206e-05
};
\label{plot:273aed8}
\addlegendentry{\footnotesize AED-8 BP-8};

\addplot[color=orange,line width = 1pt, solid, mark=x, mark options={solid}]
table[col sep=comma]{
0.00, 1.000e+00
0.50, 1.000e+00
1.00, 9.924e-01
1.50, 9.353e-01
2.00, 7.335e-01
2.50, 3.629e-01
3.00, 1.186e-01
3.50, 1.852e-02
4.00, 1.549e-03
4.50, 6.974e-05
};
\label{plot:273mbbp8}
\addlegendentry{\footnotesize MBBP-8 BP-8};

\addplot[color=black,dotted,line width = 1pt, mark=none, mark options={solid}]
table[col sep=comma]{
0.0,0.0014714285714285714
0.5,0.014
1.0,0.0375
1.5, 0.039
2.0,0.015625
2.5,0.0019615386
3.0,9.689922480620155e-05
};
\label{plot:273bp8}
\addlegendentry{\footnotesize ML lower bound};

\end{axis}

\end{tikzpicture}
    \caption{\footnotesize \Ac{BLER} performance comparison for the $(273,191)$ PG \ac{LDPC} Code.}
    \label{fig:273_cw}
\end{figure}

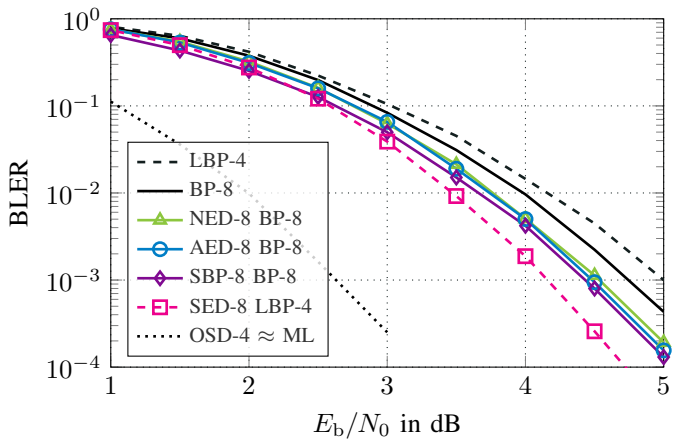
\begin{figure}
    \centering
    \begin{tikzpicture}
\begin{axis}[
width=\linewidth,
height=.7\linewidth,
grid style={dotted,anthrazit},
xmajorgrids,
yminorticks=true,
ymajorgrids,
legend columns=1,
legend pos=south west,   
legend cell align={left},
legend style={fill,fill opacity=0.8},
xlabel={$E_\mathrm{b}/N_0$ in dB},
ylabel={BLER},
legend image post style={mark indices={}},%
ymode=log,
mark size=2.5pt,
xmin=1,
xmax=5,
ymin=1e-4,
ymax=1
]

\addplot[anthrazit, dashed, mark options={solid},mark=None, line width=1.0pt, mark size= 2pt] 
table[col sep=comma]{
0.0,0.983
0.5,0.927
1.0,0.809
1.5,0.637
2.0,0.418
2.5,0.223
3.0,0.105
3.5,0.0452
4.0,0.0145
4.5,0.00447
5.0,0.001
};
\label{plot:132_LBP_4}
\addlegendentry{\footnotesize LBP-4};

\addplot[black, mark options={solid},mark=None, line width=1.0pt] 
table[col sep=comma]{
0.0,0.979
0.5,0.904
1.0,0.778
1.5,0.597
2.0,0.377
2.5,0.198
3.0,0.0834
3.5,0.0311
4.0,0.00963
4.5,0.00223
5.0,0.000432
};
\label{plot:132_FBP_8}
\addlegendentry{\footnotesize BP-8};

\addplot[apfelgruen, mark options={solid},mark=triangle, line width=1.0pt] 
table[col sep=comma]{
0.0,0.971
0.5,0.908
1.0,0.754
1.5,0.548
2.0,0.323
2.5,0.161
3.0,0.0626
3.5,0.0213
4.0,0.00506
4.5,0.00113
5.0,0.000192
};
\label{plot:132_NED_8}
\addlegendentry{\footnotesize NED-8 BP-8};	

\addplot[color=mittelblau,line width = 1pt, solid, mark=o, mark options={solid}]
table[col sep=comma]{
0.00, 9.703e-01
0.50, 8.754e-01
1.00, 7.588e-01
1.50, 5.344e-01
2.00, 3.112e-01
2.50, 1.593e-01
3.00, 6.557e-02
3.50, 1.897e-02
4.00, 5.019e-03
4.50, 9.444e-04
5.00, 1.557e-04
};
\label{plot:5guaed}
\addlegendentry{\footnotesize AED-8 BP-8};

\addplot[color=lila,line width = 1pt, solid, mark=diamond, mark options={solid}]
table[col sep=comma]{
0.00, 9.187e-01
0.50, 8.196e-01
1.00, 6.514e-01
1.50, 4.322e-01
2.00, 2.525e-01
2.50, 1.265e-01
3.00, 4.981e-02
3.50, 1.506e-02
4.00, 4.206e-03
4.50, 7.992e-04
5.00, 1.305e-04
};
\label{plot:132sbp8}
\addlegendentry{\footnotesize SBP-8 BP-8};

\addplot[magenta, dashed, mark options={solid},mark=square, line width=1.0pt] 
table[col sep=comma]{
0.0,0.965
0.5,0.892
1.0,0.741
1.5,0.498
2.0,0.276
2.5,0.121
3.0,0.0389
3.5,0.00922
4.0,0.00188
4.5,0.000258
5.0,3.09e-05
};
\label{plot:132_SED_8}
\addlegendentry{\footnotesize SED-8 LBP-4};

\addplot[color=black,line width = 1pt, dotted,mark=none, mark size=2.5pt, mark options={solid}]
table[col sep=comma]{
1.00, 1.120e-01
1.50, 3.609e-02
2.00, 9.891e-03
2.50, 1.623e-03
3.00, 2.514e-04
};
\label{plot:5g132_osd}
\addlegendentry{\footnotesize OSD-4 $\approx$ ML};

\end{axis}

\end{tikzpicture}
    \caption{\footnotesize \Ac{BLER} performance comparison for the $(132,66)$ 5G \ac{LDPC} code.}
    \label{fig:132_cw}
\end{figure}

Fig.~\ref{fig:63_cw} shows the \ac{BLER} for several decoders for the low-rate $(63,6)$ code.
For simulation, we used a systematic parity-check matrix $\Hm_\mathrm{sys}$ and a parity-check matrix containing random minimum weight checks.
In the results we only show the better of the two for the ensembles, respectively; indicated by $\Hm_\mathrm{sys}$ in the legend.
Notably, the ensembles deforming or transforming decoding regions (\ac{SED} and \ac{AED}) perform best with the systematic matrix.
In contrast, the ensembles adding noise (translating the observation) (\ac{SBP} and \ac{NED}) prefer the random matrix.
Only \ac{AED} achieves near \ac{ML} performance. Note that \ac{AED} uses for simplicity only the cyclic subgroup of the full automorphism group of the code, which is the general linear group \cite{macwilliams77}.

Fig.~\ref{fig:273_cw} demonstrates the performance of the ensembles on the $(273,191)$ \ac{PG} \ac{LDPC} code, i.e., another cyclic, thus, highly symmetrical code.
Again, we observe that the ensembles perform best that leverage the code structure (\ac{AED}, \ac{MBBP}).
Similarly, the code-agnostic \ac{SED} performs the second best, while the translation-based ensembles perform the worst.

Furthermore, Fig.~\ref{fig:132_cw} depicts the \ac{BLER} for the 5G \ac{LDPC} code with parameters $(132,66)$.
The only useful code property for the ensembles is that it is quasi-cyclic.
Note that \ac{MBBP} cannot be shown here, since the minimum weight checks are unknown.
The best performing ensemble is the \ac{SED}.
In contrast to the highly symmetric codes before, here, the \ac{AED} gain is relatively small and comparable to the translation-based ensembles.
The reason is the interplay between constituent decoder symmetries and code symmetries.
Remember that \ac{BP} is invariant to \ac{QC} permutations. 
Thus, the diversity comes only from a small deformation (deletion of 3 checks) resulting in small diversity.
Further, we observe that for all codes \ac{SBP} outperforms \ac{NED}.
This leads to the conclusion that the directed injection of noise (saturation of the least reliable positions) seems superior to undirected addition of noise, independent of the reliability,  as in \ac{NED}.

\subsection{Ensemble Gain}

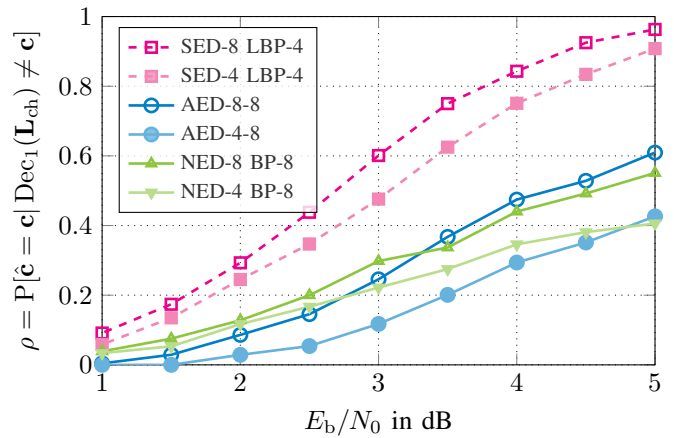
\begin{figure}
    \centering
    \begin{tikzpicture}
\begin{axis}[
width=\linewidth,
height=.7\linewidth,
grid style={dotted,anthrazit},
xmajorgrids,
yminorticks=true,
ymajorgrids,
legend columns=1,
legend pos=north west,   
legend cell align={left},
legend style={fill,fill opacity=0.8},
xlabel={$E_\mathrm{b}/N_0$ in dB},
ylabel={$\rho = \operatorname{P}[\hat{\cv}=\cv|\operatorname{Dec}_1(\Lm_\mathrm{ch})\neq\cv]$},
legend image post style={mark indices={}},%
mark size=2.5pt,
xmin=1,
xmax=5,
ymin=0,
ymax=1
]

\addplot[magenta, mark options={solid},mark=square, dashed, line width=1.0pt, mark size= 2pt] 
table[col sep=comma]{
0.0,0.0103
0.5,0.0299
1.0,0.0914
1.5,0.174
2.0,0.293
2.5,0.438
3.0,0.601
3.5,0.75
4.0,0.843
4.5,0.925
5.0,0.963
};
\label{plot:132_SED_8_P}
\addlegendentry{\footnotesize SED-8 LBP-4};

\addplot[magenta!60, dashed, mark options={solid},mark=square*, line width=1.0pt, mark size= 2pt] 
table[col sep=comma]{
0.0,0.00663
0.5,0.0225
1.0,0.0585
1.5,0.135
2.0,0.245
2.5,0.347
3.0,0.476
3.5,0.625
4.0,0.751
4.5,0.834
5.0,0.908
};
\label{plot:132_SED_4}
\addlegendentry{\footnotesize SED-4 LBP-4};

\addplot[color=mittelblau,line width = 1pt, solid, mark=o, mark options={solid}]
table[col sep=comma]{
0.0 , 0.002863864171013586
0.5 , 0.00727352876350007
1.0 , 0.004212743549236508
1.5 , 0.029013539651837505
2.0 , 0.08605423353624797
2.5 , 0.14557967178401288
3.0 , 0.24571224392567903
3.5 , 0.36750238019676296
4.0 , 0.47456704062738264
4.5 , 0.5283822138126774
5.0 , 0.6093906093906094
};
\label{plot:5guaedp}
\addlegendentry{\footnotesize AED-8-8};

\addplot[color=mittelblau!60,line width = 1pt, solid, mark=*, mark options={solid}]
table[col sep=comma]{
0.0 , 0
0.5 , 0
1.0 , 0
1.5 , 0
2.0 , 0.02877697841726623
2.5 , 0.05399682371625203
3.0 , 0.11755597903763704
3.5 , 0.20088860679149478
4.0 , 0.2938677703953818
4.5 , 0.35146641438032167
5.0 , 0.4263236763236763
};
\label{plot:5guaed4p}
\addlegendentry{\footnotesize AED-4-8};

\addplot[apfelgruen, solid, mark options={solid},mark=triangle, line width=1.0pt, mark size= 2pt] 
table[col sep=comma]{
0.0,0.00257
0.5,0.0157
1.0,0.0389
1.5,0.0751
2.0,0.128
2.5,0.2
3.0,0.298
3.5,0.337
4.0,0.44
4.5,0.492
5.0,0.551
};
\label{plot:132_NED_8_P}
\addlegendentry{\footnotesize NED-8 BP-8};	

\addplot[apfelgruen!60, solid, mark options={solid, rotate=180},mark=triangle*, line width=1.0pt, mark size= 2pt] 
table[col sep=comma]{
0.0,0.00257
0.5,0.0142
1.0,0.0337
1.5,0.0533
2.0,0.117
2.5,0.167
3.0,0.222
3.5,0.275
4.0,0.346
4.5,0.381
5.0,0.406
};
\label{plot:132_SED_4_P}
\addlegendentry{\footnotesize NED-4 BP-8};

\end{axis}

\end{tikzpicture}
    \caption{\footnotesize Ensemble recovery probability $\operatorname{P}[\hat{\cv}=\cv|\operatorname{dec}_0(\yv)\neq\cv]$ for the (132,66) 5G \ac{LDPC} code for different ensemble decoding algorithms, i.e., the probability that an ensemble will decode correctly in a case where a non-ensemble decoder would fail.}
    \label{fig:132_p}
\end{figure}

To better grasp the gain provided by the ensemble, we analyze the probability that the ensemble successfully recovers the correct codeword in the cases that the first decoder (or in general, any single decoder) fails.
Mathematically, this \emph{ensemble recovery probability} can be written as
\begin{equation}
    \rho = \operatorname{P}[\hat{\cv}=\cv|\operatorname{Dec}_1(\Lm_\mathrm{ch})\neq\cv].
\end{equation}
Intuitively, $\rho$ can be approximated by a ratio of the \acp{BLER} of the ensemble decoder and the single decoder
\begin{equation}
    \rho \approx 1 - \frac{\operatorname{P}[\operatorname{Dec}_\mathrm{E}(\Lm_\mathrm{ch})\neq\cv]}{\operatorname{P}[\operatorname{Dec}_1(\Lm_\mathrm{ch})\neq\cv]}.
\end{equation}
It is only an approximation, as \ac{ML} errors are not considered. However, it is accurate if the performance is sufficiently far away from \ac{ML}.
Fig.~\ref{fig:132_p} shows the value of the ensemble recovery probability $\rho$ with respect to $E_\mathrm{b}/N_0$ for the $(132,66)$ 5G \ac{LDPC} code for \ac{SED}, \ac{AED} and \ac{NED} with ensemble sizes $M=4$ and $M=8$, respectively.
Note that for \ac{SBP}, this metric does not make sense, as the first decoder path already uses a highly distorted version of $\Lm_\mathrm{ch}$.
As can be seen, for all ensembles, $\rho$ increases with the \ac{SNR}, i.e., the ensemble gain becomes more prominent for lower error rates.
Moreover, the difference in $\rho$ from ensemble size $4$ to $8$ is largest for \ac{AED}, indicating that it profits the most from increasing the ensemble size. However, in the particular case for the 5G \ac{LDPC} code, \ac{SED} is the only one that achieves a value of $\rho$ close to one in the regarded \ac{SNR} range.

\section{Conclusion and Outlook}
In this paper, we compared numerous ensemble decoding schemes qualitatively and quantitatively.
We showcased different mechanisms that can be leveraged to form an ensemble, i.e., transformation of the observation (\ac{AED}), transformation of the decoding region (\ac{MBBP}), deformation of the decoding region (\ac{SED} and \ac{MBBP}), and translation the observation (\ac{SBP} and \ac{NED}). 
\begin{table}[htb]
    \caption{\footnotesize Requirements and qualitative comparison of the considered ensemble decoding methods.}
    \centering
    \setlength{\tabcolsep}{3pt}
    \begin{tabular}{c|c|c|c}
        Decoder & Code Requirements & Decoder Requirements & Gain \\
        \hline
        MBBP &  $\gg m$ Min. Weight Checks & - & ++ \\
        AED & Code Automorphisms & Non-Equivariance & ++\\
        SED & - & Layered Decoder & ++ \\
        NED, SBP & - & - & + \\
    \end{tabular}
    \label{tab:comp}
\end{table}

All ensembles improve the error-rate performance compared to \ac{BP} decoding and have a similar complexity at the same ensemble size.
However, each comes with its own benefits and limitations.
In particular, the high-gain ensembles \ac{MBBP}, \ac{AED} and \ac{SED} pose constraints on the used codes and/or decoding algorithms.
These requirements are summarized in Tab.~\ref{tab:comp}.
\Ac{MBBP} requires the access to a large number of low weight parity-checks to construct suitable parity-check matrices, which are usually found only in algebraic constructions.
\Ac{AED} requires the code to have more symmetries (automorphism) than the decoder, and hence, is also best suited for algebraicly constructed codes.
\Ac{SED} in contrast only requires the use of a decoder that enables scheduling, such as a layered decoder. This comes at the cost of an increased latency due to the sequential rather than parallel processing of each iteration.
In contrast, the translation-based ensembles are applicable for any code and any constituent decoder, however, at a smaller error-rate performance gain.
Comparing the translation-base ensembles, we observed that directed noise, instead of undirected noise, leads to superior performance.

In total, we can conclude that code structure-based ensembles, like \ac{AED}, perform the best for codes with large symmetry groups, while for codes without non-equivariant symmetries, \ac{SED} is the best choice.

\appendices
\section{} \label{sec:appendix}
Only rarely the \ac{BP} decoder converges to a valid, but incorrect codeword.
This is a so-called an undetected error.
Fig.~\ref{fig:132_cw_uer} shows the \ac{UER} and the \ac{BLER} for several codes for $8$ and $32$ flooding decoding iterations.
Note that the larger the number of decoding iterations, the larger the \ac{UER}.

\begin{figure}[hb]
    \centering
    \begin{tikzpicture}
\begin{axis}[
width=\linewidth,
height=.7\linewidth,
grid style={dotted,anthrazit},
xmajorgrids,
yminorticks=true,
ymajorgrids,
legend columns=1,
legend pos=south west,   
legend cell align={left},
legend style={fill,fill opacity=0.8},
xlabel={$E_\mathrm{b}/N_0$ in dB},
ylabel={BLER/UER},
legend image post style={mark indices={}},%
ymode=log,
mark size=2.5pt,
xmin=1,
xmax=4,
ymin=1e-7,
ymax=1,
legend entries={BLER,
                UER,
                BP-8,
                BP-32},
]
\addlegendimage{no markers,black}
\addlegendimage{no markers,black,dashed}
\addlegendimage{only marks, mark=square}
\addlegendimage{only marks, mark=o}

\addplot[rot, mark options={solid},mark=square, line width=1.0pt] 
table[col sep=comma]{
0.0,0.975
0.5,0.911
1.0,0.778
1.5,0.582
2.0,0.37
2.5,0.198
3.0,0.088
3.5,0.032
4.0,0.00946
};
\label{plot:132_FBP_8_uer_bler}

\addplot[rot, dashed,mark options={solid},mark=square, line width=1.0pt] 
table[col sep=comma]{
0.0,2.18e-05
0.5,6.4e-05
1.0,0.000113
1.5,0.000158
2.0,0.00012
2.5,6.74e-05
3.0,3.21e-05
3.5,1.07e-05
4.0,2.57e-06
};
\label{plot:132_FBP_8_uer}

\addplot[rot, mark options={solid},mark=o, line width=1.0pt] 
table[col sep=comma]{
0.0,0.857
0.5,0.7
1.0,0.495
1.5,0.296
2.0,0.145
2.5,0.0569
3.0,0.0182
3.5,0.00465
4.0,0.000974
};
\label{plot:132_FBP_32_uer_bler}

\addplot[rot,dashed, mark options={solid},mark=o, line width=1.0pt] 
table[col sep=comma]{
0.0,0.00169
0.5,0.00245
1.0,0.00255
1.5,0.00175
2.0,0.00112
2.5,0.000359
3.0,0.000132
3.5,3.35e-05
4.0,8.13e-06
};
\label{plot:132_FBP_32_uer}
	
\addplot[mittelblau, mark options={solid},mark=o, line width=1.0pt] 
table[col sep=comma]{
2.0,0.795
2.5,0.48
3.0,0.185
3.5,0.0426
};
\label{plot:273_FBP_32_bler}

\addplot[mittelblau,dashed, mark options={solid},mark=o, line width=1.0pt] 
table[col sep=comma]{
2.0,2.22e-05
2.5,3.46e-05
3.0,1.87e-05
3.5,4.42e-06
};
\label{plot:273_FBP_32_uer}

\addplot[mittelblau, mark options={solid},mark=square, line width=1.0pt] 
table[col sep=comma]{
2.0,0.867
2.5,0.585
3.0,0.269
3.5,0.08322
};
\label{plot:132_FBP_8_bler}

\addplot[mittelblau,dashed, mark options={solid},mark=square, line width=1.0pt] 
table[col sep=comma]{
2.0,1.04e-06
2.5,2.27e-06
3.0,2.12e-06
3.5,6.938e-07
};
\label{plot:273_FBP_8_uer}

\draw[line width=1.0pt] (axis cs: 2.85, 2.2e-1) arc(-140:160:0.05cm and 0.15cm);
\draw[line width=1.0pt] (axis cs:2.865,1.9e-1) -- (axis cs: 2.865, 8e-3);

\draw[line width=1.0pt] (axis cs: 2.83, 1.5e-6) arc(-120:120:0.15cm and 0.45cm);
\draw[line width=1.0pt] (axis cs:2.865,2.75e-5) -- (axis cs: 2.865, 8e-3);

\node[rectangle, align=center, inner sep=1pt, fill=white] at (axis cs: 2.865, 2e-3){\footnotesize PG LDPC\\$(273,191)$};

\draw[line width=1.0pt] (axis cs: 1.85, 1.7e-1) arc(-140:130:0.05cm and 0.2cm);
\draw[line width=1.0pt] (axis cs:1.865,1.5e-1) -- (axis cs: 1.865, 8e-3);

\draw[line width=1.0pt] (axis cs: 1.86, 1.1e-4) arc(-120:120:0.15cm and 0.41cm);
\draw[line width=1.0pt] (axis cs:1.892,1.7e-3) -- (axis cs: 1.892, 8e-3);

\node[rectangle, align=center, inner sep=1pt, fill=white] at (axis cs: 1.865, 1.3e-2){\footnotesize 5G LDPC\\$(132,66)$};
\end{axis}

\end{tikzpicture}
    \caption{\footnotesize \Ac{BLER} and \ac{UER} performance comparison for the $(132,66)$ 5G \ac{LDPC} code and the $(273,191)$ \ac{PG} \ac{LDPC} code.}
    \label{fig:132_cw_uer}
\end{figure}
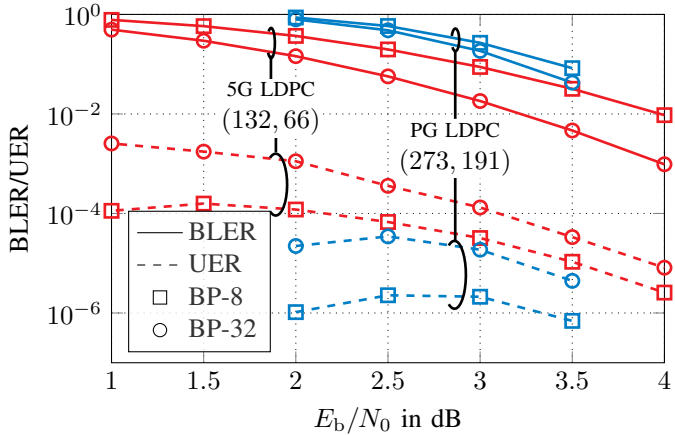

\bibliographystyle{IEEEtran}
\bibliography{references.bib}
\end{NoHyper}
\end{document}